**Chronic stress may disrupt covariant fluctuations of vitamin D and cortisol plasma levels in pregnant sheep during the last trimester: a preliminary report**

C. Wakefield[1], B. Janoschek[1], Y. Frank[1], F. Karp[2], N. Reyes[3], J. Schulkin[1], M.G. Frasch[1]

[1] Dept. of Obstetrics and Gynecology, University of Washington, Seattle, WA

[2] Depts. of Pharmacy and Bioengineering, University of Washington, Seattle, WA

[3] Dept. of Comparative Medicine, University of Washington, Seattle, WA

Address of correspondence:

Martin G. Frasch

Department of Obstetrics and Gynecology

University of Washington

1959 NE Pacific St

Box 356460

Seattle, WA 98195

Phone: +1-206-543-5892

Fax: +1-206-543-3915

Email: mfrasch@uw.edu

**Abstract**

Psychosocial stress during pregnancy is a known contributor to preterm birth, but also has been increasingly appreciated as an in utero insult acting long-term on prenatal and postnatal neurodevelopmental trajectories. These events impact many information molecules, including both vitamin D and cortisol. Both have been linked to low birth premature babies. Cortisol tends to be further elevated in women, while vitamin D tends to be decreased from their normal levels during pregnancy. One facilitates labor in part by elevating placental CRH, the other by limiting CRH in placental tissue. Both are linked to managing adversity. Studies in large animal models with high resemblance to human physiology are sparse to model the changes induced by such stress exposure. Using an established pregnant sheep model of stress during human development, here we focused on measuring the changes in maternal Vitamin D and cortisol responses due to chronic inescapable stress mimicking daily challenges in the last trimester of human pregnancy. The present pilot data show that chronic maternal stress during pregnancy results in endocrine and metabolic chronic habituation paralleled by sensitization to acute stress challenges. Chronic stress appears to disrupt a physiological relationship between oscillations of vitamin D and cortisol. These speculations need to be explored in future studies.

**Introduction**

Psychosocial stress during pregnancy is a known contributor to preterm birth(Shapiro et al., 2013), but also has been increasingly appreciated as an *in utero* insult acting long-term on prenatal and postnatal neurodevelopmental trajectories.(Frasch et al., 2018)

These events impact many information molecules, including both vitamin D and cortisol.(Mohamed et al., 2018; Schulkin, 2017; Wang et al., 2018) Both have been linked to low birth premature babies; cortisol tends to be further elevated in women; vitamin D tends to be decreased from their normal levels during pregnancy.(Mohamed et al., 2018; Schulkin, 2017; Wang et al., 2018) One facilitates labor in part by elevating placental CRH (Wang et al., 2018); the other by limiting CRH in placental tissue. Both are linked to managing adversity.(Asok et al., 2016; Schulkin, 2017)

Studies in large animal models with high resemblance to human physiology are sparse to model the changes induced by such stress exposure. Using an established pregnant sheep model of stress during human development (Rakers et al., 2013), here we focused on measuring the changes in maternal Vitamin D and cortisol responses due to chronic inescapable stress mimicking daily challenges in human pregnancy's last trimester.

**Materials and Methods**

The study was approved by the Institutional Animal care and Use Committee (IACUC) of the University of Washington (protocol number 4403-01).

Five (three with twins, two with singletons) pregnant sheep were delivered to the University of Washington animal facility at 82 days of gestation, kept in quarantine for 18 days to confirm the Q fever negative status, and chronically instrumented with maternal jugular catheters at 111±2 days of gestation (dGA, term 145 days, human week 30 equivalent) and weighed 65±2 kg. The first ewe was used for surgery pilot and was not enrolled in the study. Following 3-4 days of recovery, the remaining four animals continued in the study. In the course of four consecutive weeks the animals were subjected to three hours lasting irregularly occurring complete isolations, twice weekly and at least 48 hours apart, *i.e.*, a total of eight isolations. Camera surveillance was deployed to ensure overall animal well-being. The procedures were well tolerated. At the end of the four weeks period, the animals were euthanized as described before.(Burns et al., 2015)

For isolation, the stressed group's pregnant sheep was brought into a separate room, devoid of any stimulation. Immediately before and at the end of each isolation, a maternal jugular vein 4 mL sample was obtained, immediately spun down (4 C, 4,000 rpm, 4 min), plasma frozen in two aliquots. The controls were sampled at the matched time points, but not isolated. A total of 56 samples were analyzed for Vitamin D and cortisol concentrations using commercially available assays in the Phoenix Lab (4338 Harbour Pointe Boulevard S.W., Mukilteo, WA, 98275) and the Michigan St. Veterinary Diagnostic Laboratory (4125 Beaumont Road, Lansing, MI, 48910-8104). Cortisol serum levels were quantified using a solid-phase competitive enzyme chemiluminescent immunoassay. Vitamin D serum levels were determined through an ELISA assay for the metabolically active form of the hormone, calcitriol.

Statistical analysis was conducted in two control animals and two chronically stressed animals using Generalized Estimating Equations (GEE) modeling with scale weight adjustment for "pre" and "post" isolation measurements in control and stressed animals. The model terms included the main term group (control or stressed) and the interaction term group*isolation number to measure both, the overall impact of chronic inescapable stress as well as the cumulative effect of isolations. Adjusting for "pre" and "post" time points allowed us to also account specifically for the chronic ("pre") and acute ("post") effects of stress. For correlation analyses, Pearson statistics was used. Linear multivariate regression analysis was performed in Exploratory (R - based data science application). Statistical significance was assumed at p-value < 0.05. The data has been deposited on [GitHub](GitHub).

**Results**

Animal health characteristics are summarized in Fig. 1A. While we observed no overall group effect of stress on $pO_2$ and $pCO_2$ levels (group terms p=0.351 and p=0.56, respectively), accounting for repetitive isolations there was a significant effect (group*isolation number interaction term p<0.001 for both). For lactate, both group and interaction terms (group*isolation number) were significant at p=0.004 and p<0.001, respectively.

Cortisol levels were different on both group and interaction term levels (p=0.031 and p<0.001, respectively). For Vitamin D, group term was not significant (p=0.237), while the interaction term group*isolation number was (p<0.001).

Next, we studied the relationship between Vitamin D and Cortisol over the course of isolations (Fig. 1B) using linear multivariate regression analysis. Since the direction of the relationship was not clear *a priori*, we performed the analysis twice. Once, predicting Vitamin D from cortisol, isolation sequence and it being a measurement pre or post isolation. In control animals we found adjusted $R^2=0.427$ (p=0.0015), while for stressed animals $R^2=0.027$ (p=0.308). In contrast, predicting cortisol from Vitamin D, also accounting for isolation sequence and pre/post isolation measurement, rendered adjusted $R^2=0.205$ (p=0.045) in controls and $R^2=-0.029$ (p=0.545) in stressed animals. These findings suggest a directional relationship between Vitamin D and cortisol fluctuations with cortisol driving in part the changes measured in Vitamin D concentrations in control ewes, but not in stressed ewes.

**Discussion**

The present data suggest that chronic maternal stress during pregnancy results in endocrine and metabolic chronic habituation paralleled by sensitization to acute stress challenges. While fetal adaptations to chronic maternal stress have been described, this is the first demonstration of similar effects on the maternal side, in particular, the effect on vitamin D fluctuations in relation to cortisol. Chronic stress appears to disrupt a physiological relationship between oscillations of vitamin D and cortisol.

Limitations of the above findings for blood gases are that they are taken from ewe's venous compartment as is standard for chronic sampling preparation in this animal model. Still, the maturational and stress-dependent changes are of note.

The relative acute ("post" isolation measurements), but not chronic ("pre" isolation measurements), maternal hypoxemia, hypercapnia, and lactemia in the stress group resembles conceptually our finding in the human pregnant daily hassles stress cohort (Lobmaier et al., 2019) where the newborns of chronically stressed mothers had lower arterial umbilical cord blood $pO_2$ at 18 vs 21. Similarly, Schwab's team, who established this ovine chronic stress paradigm, showed that chronic stress results in chronic hypoxemia likely due to a reduction in uterine blood flow due to chronic sympathetic activation of uterine vasculature.(Rakers et al., 2013)) The present findings put the fetal physiological adaptations to chronic maternal stress in new perspective suggesting that in addition to changes in uterine blood flow, systemic homeokinetic shifts may occur on the maternal side in response to chronic inescapable stress. Such shifts can synergize with the changes on the level of uterine-placental blood flow rendering the fetus of stressed mothers relatively more hypoxic.

Recent evidence suggests that vitamin D inhibits CRH and other pro-labor genes in human placenta.(Wang et al., 2018) Our findings suggest that chronic stress during pregnancy reverses the direct relationship between vitamin D and cortisol levels. Future studies may explore further the changes the placenta tissues to corroborate this observation.

Recent evidence suggests that vitamin D levels may be separate from hypothalamic-pituitary-adrenal (HPA) axis related events.(Ayers et al., 2018) CRH and glucocorticoid levels, perhaps extrahypothalamic expression might be altered. The present results are consistent with this observation.

If vitamin D couples differently with CRH under stress, this may be clinically important to know for pregnant women and care providers: vitamin D supplementation under stressful conditions of pregnancy may have different effects on a key molecule timing the birth, the CRH. These speculations need to be explored in future studies.

**Word count: 1577**

**Acknowledgments**

We gratefully acknowledge the team of the Department of Comparative Medicine at UW. We thank in particular Gary Fye.

**Declaration of interest statement**

The authors declare that no conflicts of interest exist.

**Figure captions.**

**Figure 1. A.** Blood gas, lactate, cortisol, and vitamin D responses to chronic stress by intermittent isolations (stress group) compared to sham control animals, plotted separately for pre- and post-isolation time points. Gestational age (dGA) is shown on each subsequent isolation day, from isolation #1 (dGA 112) to isolation #8 (dGA 137).

**B.** Individual time points of Cortisol and Vitamin D fluctuations over the period of isolations (pre-isolation, top) and post-isolation (bottom) shown on X-axis (number of isolation, from first to eighth time). Y-axis is log scale cortisol and vitamin D concentrations. Note evidence of the temporal correlation between both hormone fluctuation patterns in sham control ewes, but not in stressed ewes. Natural fluctuations of cortisol are observed in sham control animals. These fluctuations appear to be entrained with vitamin D fluctuations in sham control ewes, but not in the stressed ewes.

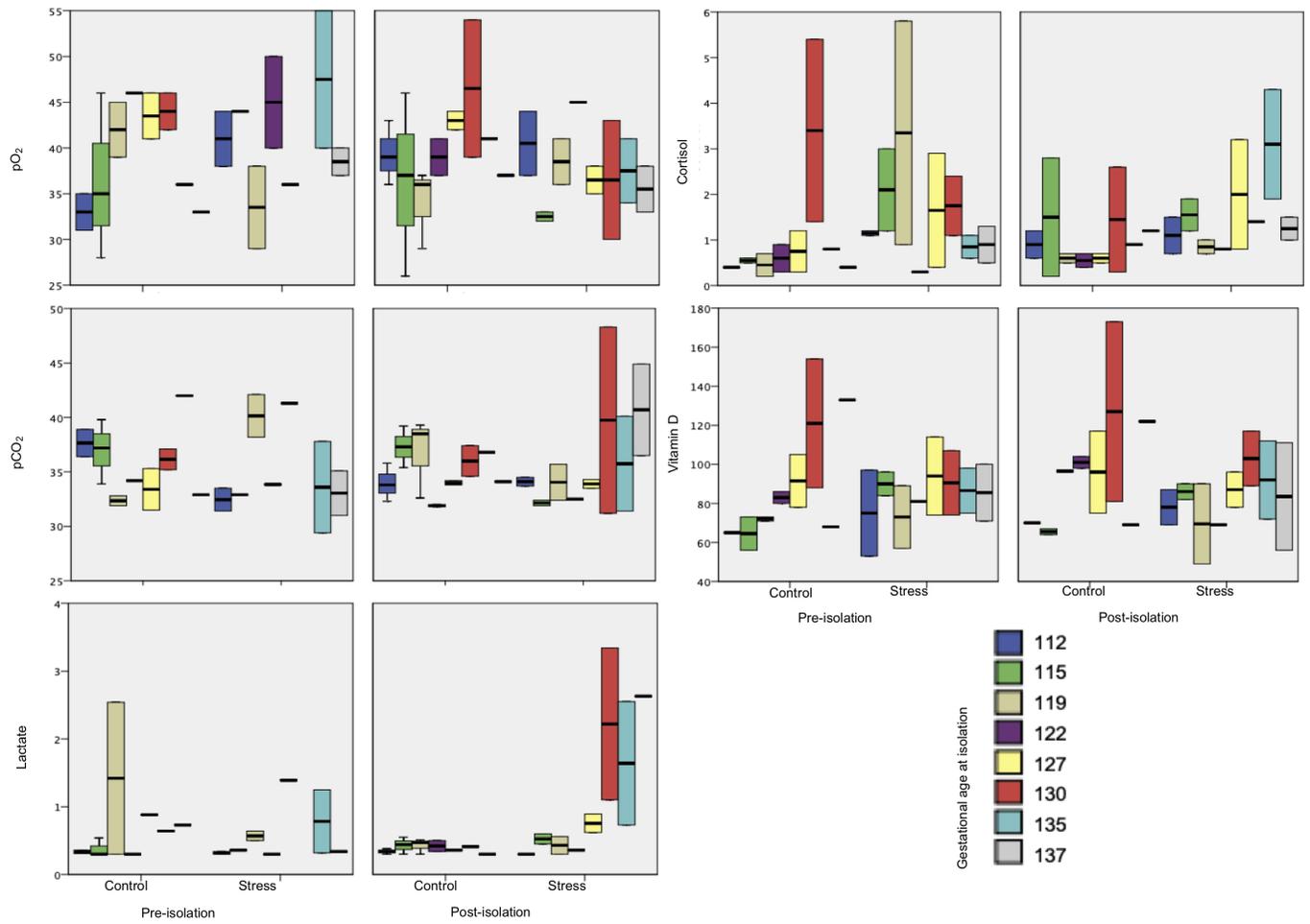

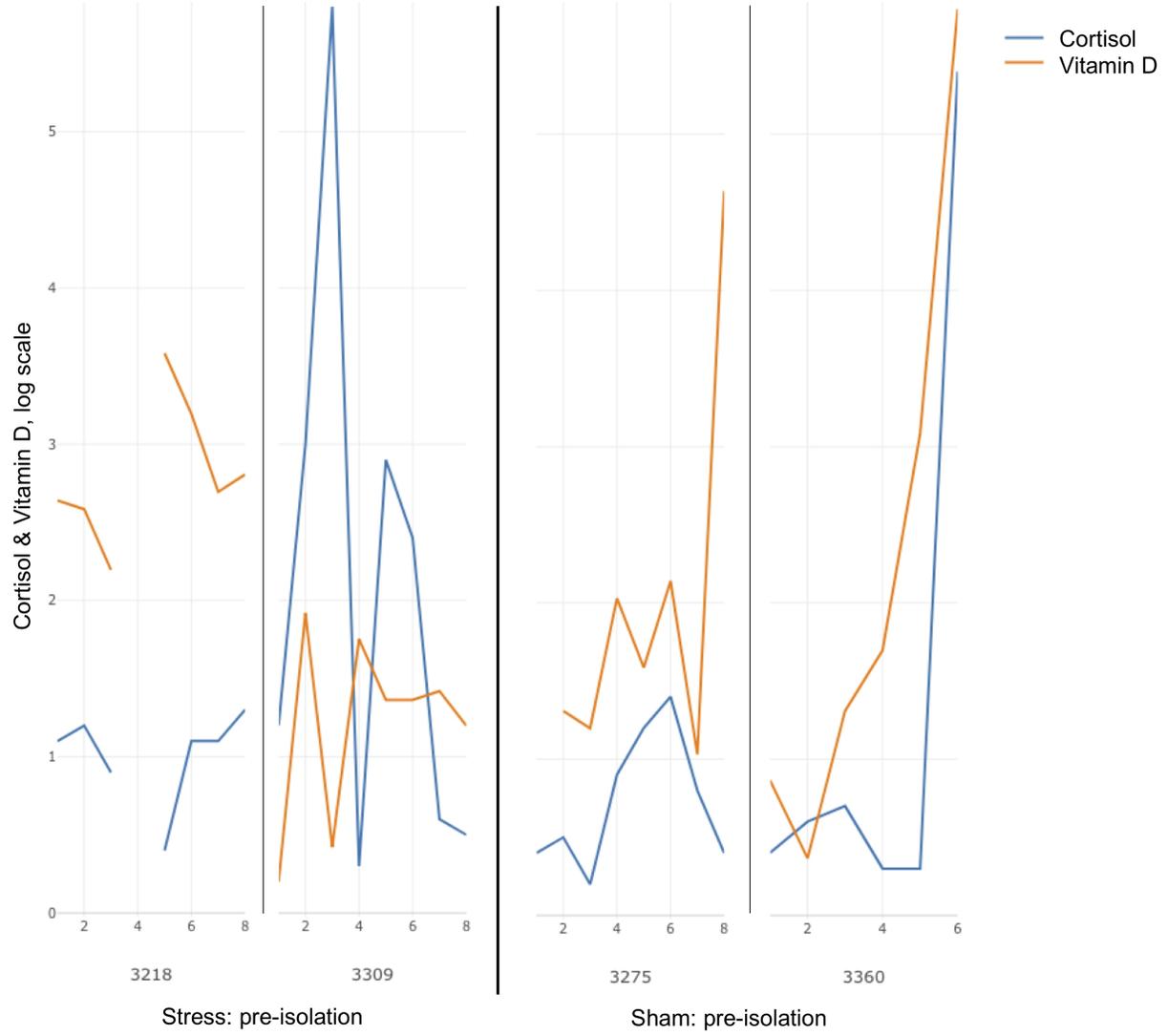

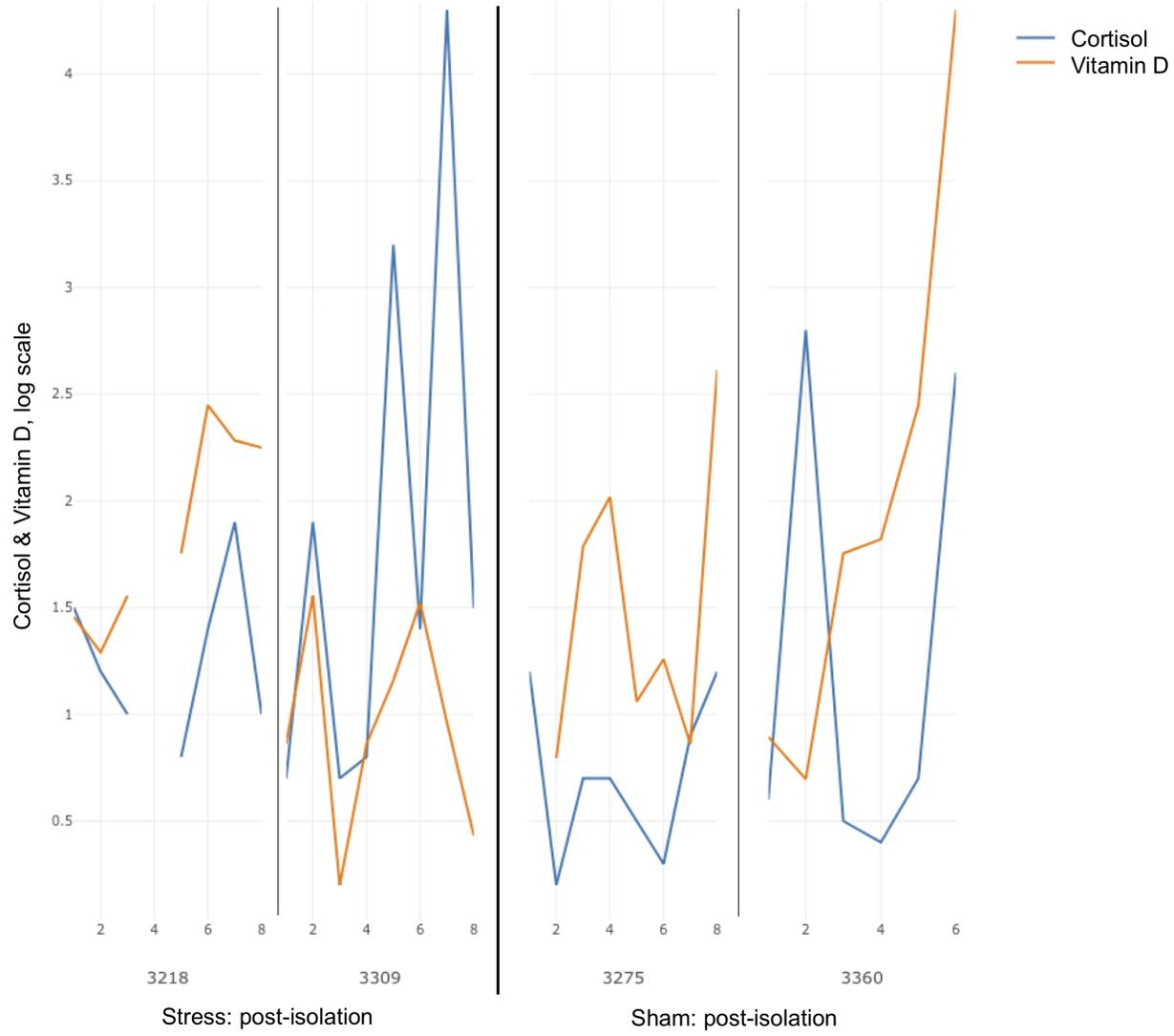